\documentclass[aps,pra,twocolumn,amsmath,amssymb,nofootinbib,showpacs,superscriptaddress]{revtex4-2}
\usepackage[english]{babel}
\usepackage{latexsym}
\usepackage{graphics}
\usepackage{graphicx}
\usepackage{epsfig}
\usepackage{color}
\usepackage{bm}
\usepackage{amsmath}
\usepackage{amssymb}
\usepackage{amsthm}
\usepackage{dcolumn}
\usepackage{bbm}
\usepackage{float}
\usepackage{hyperref}
\usepackage{color}
\usepackage{epstopdf}
\usepackage{cleveref}
\usepackage[svgnames]{xcolor}
\usepackage{physics}
\usepackage[sort&compress]{natbib}
\usepackage{bbold}
\usepackage{enumerate}
\usepackage{enumitem}

\hypersetup{hidelinks,colorlinks=true,allcolors=DarkBlue}

\newcommand{\BLUE}[1]{{\color{black} #1}}

\begin{document}
\preprint{APS/123-QED}

\title{Evidence for effectively constant shot complexity in the quantum approximate optimization algorithm without per-instance optimization}

\newcommand{\RQC}{Russian Quantum Center, Skolkovo, Moscow 121205, Russia}
\newcommand{\FIZTECH}{Moscow Institute of Physics and Technology, Dolgoprudny 141700, Russia}
\newcommand{\MISIS}{National University of Science and Technology ``MISIS'', Moscow 119049, Russia}
\newcommand{\STEKLOV}{Department of Mathematical Methods for Quantum Technologies, Steklov Mathematical Institute of Russian Academy of Sciences, Moscow 119991, Russia}
\newcommand{\KURCHATOV}{NRC ``Kurchatov Institute'', Moscow 123182, Russia}

\author{Andrey~Yu.~Chernyavskiy}
\affiliation{\RQC}
\affiliation{\KURCHATOV}

\author{Denis~A.~Kulikov}
\affiliation{\RQC}
\affiliation{\FIZTECH}

\author{Boris~I.~Bantysh}
\affiliation{\RQC}
\affiliation{\KURCHATOV}

\author{Yurii~I.~Bogdanov}
\affiliation{\RQC}
\affiliation{\KURCHATOV}

\author{Aleksey~K. Fedorov}
\affiliation{\RQC}
\affiliation{\MISIS}

\author{Evgeniy~O. Kiktenko}
\affiliation{\RQC}
\affiliation{\MISIS}
\affiliation{\STEKLOV}

\begin{abstract}
We study a modified fixed-point version of the Quantum Approximate Optimization Algorithm (fpQAOA), where parameters are trained classically on small instances and then transferred to larger problems. Our scheme combines three ingredients: (i) targeting approximate solutions via a prescribed approximation ratio (AR), (ii) scaling the circuit depth linearly with the problem size using a two-parameter sin–cos angle encoding, and (iii) normalizing QUBO Hamiltonians by their Frobenius norm. Noiseless numerical simulations (for system sizes up to 30 qubits) across a variety of random QUBO ensembles show that with these modifications the median number of quantum circuit runs (“shots”) required to achieve AR=0.95 counterintuitively decreases towards a nearly constant value as the problem size increases, while the per-shot time remains polynomial. Extrapolation of this finite‑size behavior is consistent with an effectively constant sampling complexity. Moreover, removing any single component of the scheme restores rapid growth of the required number of shots, highlighting the synergistic nature of the three modifications. These empirical findings suggest that fpQAOA, equipped with the proposed protocol, may achieve scalable approximate performance with polynomial‑depth circuits for the considered problem classes.

\end{abstract}

\maketitle

\section{Introduction}
Quantum Approximate Optimization Algorithm (QAOA)~\cite{farhi2014qaoa} is one of the most actively studied 
near-term quantum algorithms, with demonstrated applications to QUBO and related combinatorial problems of practical 
relevance~\cite{montanaro2016quantum,glover2022quantum,semenov2024transforming,fedorov2022quantum, kotil2025quantum}. 
Its feasibility has been shown on both superconducting~\cite{harrigan2021quantum,pelofske2024short} and trapped-ion 
platforms~\cite{pagano2020quantum,shaydulin2023qaoawith}. 
The prevailing strategy for implementing QAOA is a hybrid quantum--classical optimization loop~\cite{guerreschi2019qaoa, zhou2020quantum, fernandez2022study}. While promising, this approach is hindered by several notable challenges. A primary obstacle is the statistical noise inherent in finite quantum measurements, which obscures the optimization landscape. Another significant challenge is the issue of barren plateaus~\cite{larocca2025barren,mcclean2018barren,bharti2022noisy,nemkov2025barren}, which can stifle convergence. It is important to note that these limitations extend beyond QAOA and are, in fact, characteristic of variational quantum algorithms as a whole, including quantum neural networks~\cite{mcclean2018barren,bharti2022noisy}.

However, the original paper~\cite{farhi2014qaoa} proposed a different approach. The authors offered an efficient method to compute objective function on a classical computer for a particular type of QUBO problems (Max-Cut on 3-regular graphs). In this framework, the quantum computer is only required for sampling, while parameter optimization is performed classically. This approach was developed into the alternative to the hybrid-optimization, called fixed-point QAOA (fpQAOA),  
where parameters trained on small instances are transferred to larger ones \cite{brandao2018fixed}. 
Subsequent research has explored parameter transferability~\cite{galda2021transferability,lotshaw2021empirical,lotshaw2022scaling,Shaydulin2023,akshay2021parameter,chernyavskiy2023fp,pelofske2025evaluating}, 
angle schedules such as Fourier or ramp-like ansätze~\cite{zhou2020quantum,montanez2025toward,sakai2024linearly}. 
When analyzing the effect of circuit depth in fpQAOA, prior studies have primarily focused on exact optimization tasks, often under an exponential–time scaling model for the success probability within fixed-depth circuits~\cite{boulebnane2024solving,montanez2025toward, priestley2025}.
In the exact–solution context, it was observed that beyond a certain depth the performance saturates and does not improve further~\cite{shaydulin2024evidence}. However, it is important to emphasize that this approach has provided notable numerical evidence of scalability-related quantum advantage in LABS~\cite{shaydulin2024evidence} and random k-SAT~\cite{boulebnane2024solving} problems for exact solutions.
Another important performance metric for characterizing QAOA is the approximation ratio (AR)~\cite{farhi2014qaoa,larkin2022evaluation}, which has also been employed in recent QAOA modifications, including fpQAOA~\cite{golden2023numerical,montanez2025toward,jiang2025improving}.

Despite this progress, there is no consistent evidence that fpQAOA maintains stable approximation-ratio (AR) performance 
under polynomial-depth circuits; several studies even report degradation of AR with increasing $n$~\cite{sakai2024linearly,mooney2025optimization}.  
Prior work has emphasized that understanding the joint interplay of depth, problem size, and AR is crucial~\cite{he2024parameter,golden2023numerical}.

\begin{figure*}
    \centering
    \includegraphics[width=0.99\linewidth]{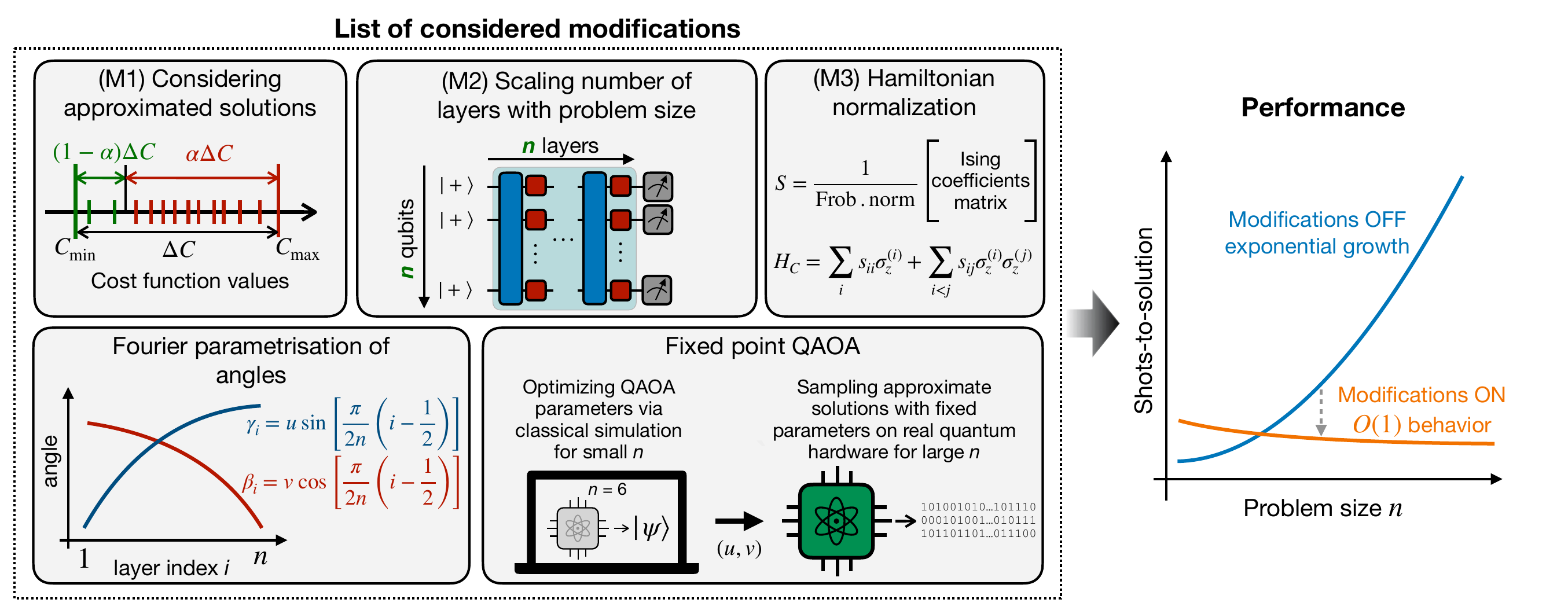}
    \caption{Schematic overview of the three modifications introduced in our fpQAOA scheme. 
    (M1) Approximate solutions are accepted based on an AR threshold $\alpha$. 
    (M2) The circuit depth is scaled with problem size ($p=n$). 
    (M3) QUBO instances are normalized by the Frobenius norm. 
    Combined with a low-dimensional Fourier encoding of angles and a fixed-point training procedure that transfers parameters from small to large instances, these ingredients change the sampling complexity from exponential scaling (blue) to an effectively constant $O(1)$ behavior (orange).
}

    \label{fig:sketch}
\end{figure*}

Here we address this challenge by constructing a concrete fpQAOA protocol combining  
(i) an AR-based objective,  
(ii) depth--size scaling $p=n$ with a two-parameter sin--cos encoding, and  
(iii) Frobenius normalization of QUBO matrices (see also Fig.~\ref{fig:sketch}).  
We show that, with these modifications, the median number of quantum circuit runs (shots) required to reach AR $\alpha=0.95$ 
becomes effectively independent of $n$.
To the best of our knowledge, this is the first demonstration of a non‑increasing median shots complexity for approximate QUBO solutions under polynomial‑depth circuits without any instance‑specific optimization process.
Notably, removing any single component of the protocol restores exponential scaling, demonstrating the synergistic nature of the method.

The paper is organized as follows.
Section~\ref{sec:preliminaries} provides the necessary background, 
Section~\ref{sec:modqaoa} outlines the proposed method,
Section~\ref{sec:performance} presents the numerical analysis and main results,
Section~\ref{sec:discussion} points the assumptions and limitations of the research, summarizes our findings and concludes the paper.

\section{Preliminaries} \label{sec:preliminaries}
\subsection{The QAOA ansatz}
Originally the QAOA~\cite{farhi2014qaoa} was proposed as a quantum algorithm designed to approximately solve combinatorial optimization problems
by preparing a parameterized quantum state. Conceptually, QAOA can be viewed as a discretized version
of adiabatic quantum evolution~\cite{farhi2000adiabatic}, where the continuous interpolation between a mixing and a problem
Hamiltonian is replaced by a finite sequence of alternating unitary operators.

Specifically, QAOA prepares the $n$-qubit state
\begin{equation} \label{eq:qaoa_state}
    \ket{\boldsymbol\beta,\boldsymbol\gamma}
    =
    \prod_{i=1}^p e^{-\imath \beta_i H_M}\, e^{-\imath \gamma_i H_C}
    \ket{+}^{\otimes n},
\end{equation}
where the unitary operators are applied sequentially layer by layer, starting from $i=1$ and proceeding up to $i=p$.
Here $H_C$ is the cost Hamiltonian encoding the objective function of the optimization problem, while $H_M$ is the mixing Hamiltonian that drives transitions between computational basis states and enables exploration of the solution space.
The parameters $\boldsymbol\beta=(\beta_1,\ldots,\beta_p)$ and $\boldsymbol\gamma=(\gamma_1,\ldots,\gamma_p)$
are variational angles, and $p$ is the circuit depth (number of layers).

In the standard formulation, the mixing Hamiltonian is chosen as
$H_M=\sum_{i=1}^n \sigma_x^{(i)}$, which promotes transitions between computational basis states,
while the cost Hamiltonian $H_C$ encodes the objective function of the optimization problem.
For quadratic unconstrained binary optimization (QUBO) problems, this objective can be written in the Ising form
\begin{equation} \label{eq:qubo_problem}
    C(z)=\sum_{i<j} s_{ij} z_i z_j + \sum_i s_{ii} z_i,\qquad z_i\in\{\pm1\},
\end{equation}
defined for spin variables $z=(z_1,\ldots,z_n)$ with real coefficients $s_{ij}$.
The corresponding operator form is obtained via the standard substitution $z_i\mapsto\sigma_z^{(i)}$.
Here $\sigma_x^{(i)}$ and $\sigma_z^{(i)}$ denote the Pauli $X$ and $Z$ operators acting on the
$i$-th qubit (and as identity on all others), and $\ket{+}=2^{-1/2}(\ket{0}+\ket{1})$ is the
$+1$ eigenstate of $\sigma_x$.

The performance of QAOA is typically quantified by the expectation value of the cost Hamiltonian,
\begin{equation}
    E(\boldsymbol\beta,\boldsymbol\gamma)
    =
    \langle\boldsymbol\beta,\boldsymbol\gamma|H_C|\boldsymbol\beta,\boldsymbol\gamma\rangle.
\end{equation}
On quantum hardware, this quantity is estimated via repeated measurements of the state
$\ket{\boldsymbol\beta,\boldsymbol\gamma}$ in the computational basis. Given
$N_{\rm samples}$ measurement outcomes
$\{z^{(1)}, \ldots, z^{(N_{\rm samples})}\}$, the expectation value is approximated as
\begin{equation} \label{eq:estimate}
    E(\boldsymbol\beta,\boldsymbol\gamma)
    \approx
    \frac{1}{N_{\rm samples}} \sum_{j=1}^{N_{\rm samples}} C(z^{(j)}),
\end{equation}
where computational-basis measurement outcomes $b_{j}^{(i)}\in\{0,1\}$ are mapped to spin variables
$z_{i}^{(j)}\in\{+1,-1\}$ via the standard transformation $z_{i}^{(j)} = 1-2{b_{i}^{(j)}}$. The statistical error of this estimator decreases as
$N_{\rm samples}^{-1/2}$.

In its standard implementation, QAOA relies on a hybrid quantum--classical optimization loop,
where the parameters $\boldsymbol\beta$ and $\boldsymbol\gamma$ are iteratively updated using classical
optimization routines. While increasing the circuit depth $p$ generally improves the quality of the obtained solutions,
it also makes the parameter optimization problem significantly more challenging.
A central challenge in this approach is the complexity of the resulting parameter landscape,
which can become highly nontrivial even for moderate circuit depths and problem sizes.
In particular, the optimization procedure may suffer from issues such as barren plateaus~\cite{mcclean2018barren}
and sensitivity to noise and sampling errors on near-term quantum devices~\cite{bharti2022noisy}.
Moreover, the cost of estimating the objective function with sufficient precision can dominate the overall
runtime, especially for large-scale instances.

We emphasize that, in practically relevant regimes, the total sampling budget accumulated over all optimization iterations involving estimates of the form~\eqref{eq:estimate} is generally expected to remain well below the size of the full configuration space, which scales as $2^n$. As a result, QAOA operates in a strongly sampling-limited regime, highlighting the importance of resource-efficient parameter optimization strategies.

\BLUE{
\subsection{Parameter transfer and fpQAOA}

To circumvent the limitations of hybrid optimization, fpQAOA replaces the instance-wise parameter search
with a parameter transfer approach~\cite{brandao2018fixed,galda2021transferability,akshay2021parameter,lotshaw2021empirical,lotshaw2022scaling,Shaydulin2023,chernyavskiy2023fp}.
The underlying idea of parameter transfer is based on the observation that the QAOA state
$\ket{\boldsymbol\beta,\boldsymbol\gamma}$ depends on a specific problem instance only through the cost Hamiltonian $H_C$,
while the variational parameters $(\boldsymbol\beta,\boldsymbol\gamma)$ control the structure of the quantum evolution in a problem-independent manner.
As a result, when the instances are drawn from the same problem class, the corresponding QAOA states adapt to changes
in the objective function through $H_C$, while the same set of parameters can still induce qualitatively similar dynamics.
This leads to the empirical observation that a single set of angles $(\boldsymbol\beta_*,\boldsymbol\gamma_*)$
can produce high-quality solutions across a wide range of instances with different cost functions,
provided that these instances share common structural properties.

The overall procedure can be naturally divided into two distinct stages: a \emph{training stage} and a \emph{sampling stage}.

In the training stage, the goal is to determine a set of universal circuit parameters
$(\boldsymbol\beta_*,\boldsymbol\gamma_*)$ that are well-suited for a given class of optimization problems.
This is achieved by optimizing the QAOA objective over a training set of relatively small QUBO instances
drawn from the target problem distribution. Importantly, this stage is performed entirely classically,
typically using statevector simulation, which allows one to get an access to $\ket{\boldsymbol\beta,\boldsymbol\gamma}$ and so to the exact value of $E(\boldsymbol\beta,\boldsymbol\gamma)$, avoiding statistical sampling noise, and
significantly simplifies the optimization procedure. From this perspective, the training process can be viewed
as finding a fixed point in the parameter space that captures the typical structure of the problem class.

In the sampling stage, the precomputed parameters $(\boldsymbol\beta_*,\boldsymbol\gamma_*)$ are directly
used to prepare the QAOA state on a quantum device for a new problem instance $C$, potentially of much larger size.
The role of the quantum processor in this stage is to generate candidate solutions
$\{z^{(1)},\ldots,z^{(M)}\}$ by sampling from the state $\ket{\boldsymbol\beta_*,\boldsymbol\gamma_*}$,
constructed with respect to the target cost function $C$, using a moderate number of samples $M$,
without any further parameter optimization.
A solution to the optimization problem can then be obtained
by selecting the sample that yields the best cost function value $C(z^{(i)})$.
As a result, the costly hybrid optimization loop is completely eliminated at the sampling stage.

In the following, we demonstrate how this scheme can be further improved by enabling the computation
of $(\boldsymbol\beta_*,\boldsymbol\gamma_*)$ using classical simulation on relatively small system sizes (e.g., $n=6$),
while simultaneously reducing the required number of samples $M$ to only a small constant in order to obtain
high-quality solutions.

}
\section{Modified fpQAOA} \label{sec:modqaoa}

To improve the transferability and scalability of the fpQAOA,
we introduce three complementary modifications to the standard training procedure: (M1) targeting approximate rather than exact solutions through the success probability at a prescribed AR, (M2) scaling the circuit depth linearly with the problem size using a two-parameter sin–cos angle encoding, and (M3) normalizing QUBO Hamiltonians by their Frobenius norm.

In the following subsections, we discuss each of these modifications in detail.
\BLUE{To gain additional insight into the structure of the learned fixed-point parameters, Appendix~\ref{app:adiabaticity} analyzes their relation to adiabatic quantum optimization and shows that the resulting fpQAOA trajectory approximately follows a low-energy path of the corresponding interpolating Hamiltonian.}

\subsection{AR-based loss function}

Common training objectives for QAOA-based optimization include the expectation value of the cost Hamiltonian~\cite{wurtz2021fixed} and the success probability of sampling the exact optimum,
\begin{equation}
    P_{\min}(\boldsymbol\beta,\boldsymbol\gamma)=
    \sum_{z:\,C(z)=c_{\min}}
    \abs{\braket{z}{\boldsymbol\beta,\boldsymbol\gamma}}^2.
\end{equation}
However, exact optimization rapidly becomes impractical as the problem size $n$ increases.
For minimization problems with optimum $c_{\min}$ and maximum value $c_{\max}$,
we therefore focus on approximate solutions characterized by the AR
$\alpha$~\cite{vazirani2001approximation}. The corresponding feasible set is defined as
\begin{equation}
    \mathcal{F}(\alpha)=
    \Bigl\{
        z :
        \frac{C(z)-c_{\min}}{c_{\max}-c_{\min}}
        \le 1-\alpha
    \Bigr\},
\end{equation}
and we introduce the associated success probability
\begin{equation}
    \label{eq:succ_prob}
    P_\alpha(\boldsymbol\beta,\boldsymbol\gamma)
    =
    \sum_{z\in\mathcal{F}(\alpha)}
    \abs{\braket{z}{\boldsymbol\beta,\boldsymbol\gamma}}^2.
\end{equation}
This quantity is used both as the training objective during the training stage and as the primary performance metric during the sampling stage. Note that ${P_{\alpha=1}=P_{\min}}$.

In particular, the parameters $\boldsymbol\beta$ and $\boldsymbol\gamma$ are optimized to maximize the worst-case value of
$P_{0.95}(\boldsymbol\beta,\boldsymbol\gamma)$ over the training set.
This choice favors parameter configurations that generalize well across different problem instances rather than overfitting to a particular example.

Several remarks are important in this context.
First, evaluating
$P_\alpha(\boldsymbol\beta,\boldsymbol\gamma)$
during training is feasible because the training stage relies on classical statevector simulation, providing direct access to the full probability distribution over computational basis states.
Second, for large problem sizes, the exact value of
$P_\alpha(\boldsymbol\beta,\boldsymbol\gamma)$
is generally unknown, similarly to the optimal cost value itself for NP-hard optimization problems.
In Sec.~\ref{sec:performance}, we nevertheless use this quantity as a benchmark metric in numerical simulations, where exact classical evaluation remains possible.

Finally, although decreasing the target AR threshold $\alpha$ generally enlarges the feasible set $\mathcal{F}(\alpha)$, potentially even exponentially with $n$ for certain problem classes, this does not imply that finding such approximate solutions becomes computationally easy.
Indeed, our numerical simulations indicate that the relative size
$\abs{\mathcal{F}(\alpha)}/2^n$
still decreases exponentially with system size for typical problem instances.
This behavior is consistent with the observed exponential scaling of brute-force search complexity and highlights that sampling high-quality approximate solutions remains a nontrivial computational task.

\subsection{Two-parameter depth--size scaling ansatz}

Recall that the standard QAOA ansatz~\eqref{eq:qaoa_state} consists of
$p$ alternating layers of cost and mixer unitaries, resulting in a total of
$2p$ variational parameters
$(\boldsymbol\beta,\boldsymbol\gamma)$.
It is well known that increasing the circuit depth $p$ for fixed problem size $n$
generally improves QAOA performance.
At the same time, keeping $p$ fixed while increasing $n$ typically leads to an
exponential degradation in success probability.
In our approach, we therefore impose the scaling relation
\begin{equation}
    p=n,
\end{equation}
which, together with the other modifications introduced in this work,
constitutes a key ingredient for maintaining a nearly constant shot complexity
with increasing system size.

In addition, rather than optimizing all QAOA angles independently,
one can represent them using a lower-dimensional functional parametrization.
A widely used approach is the Fourier parametrization~\cite{zhou2020quantum},
defined as
\begin{equation}\label{eq:fourier}
\begin{aligned}
\gamma_i &=
\sum_{k=1}^q
u_k
\sin\qty[
    \qty(k-\frac12)
    \qty(i-\frac12)
    \frac{\pi}{p}
],
\\
\beta_i &=
\sum_{k=1}^q
v_k
\cos\qty[
    \qty(k-\frac12)
    \qty(i-\frac12)
    \frac{\pi}{p}
],
\end{aligned}
\end{equation}
where
$\{u_k\}_{k=1}^q$
and
$\{v_k\}_{k=1}^q$
are $2q$ trainable parameters.
Choosing $q<p$ reduces the dimensionality of the optimization problem.

In this work, we employ the simplest nontrivial case corresponding to $q=1$.
Combined with the depth--size relation $p=n$, this yields the two-parameter schedule
\begin{equation} \label{eq:fourier_2p}
\begin{aligned}
    \gamma_i &=
    u
    \sin\qty[
        \qty(i-\frac12)
        \frac{\pi}{2n}
    ],\\
    \beta_i &=
    v
    \cos\qty[
        \qty(i-\frac12)
        \frac{\pi}{2n}
    ],
\end{aligned}     
\end{equation}
with $i=1,\dots,n$.

We note that many alternative parametrization strategies for QAOA angles are possible,
including higher-order Fourier expansions and other smooth functional encodings.
However, as demonstrated in Sec.~\ref{sec:performance},
the simple two-parameter ansatz considered here already provides highly competitive performance,
indicating that more complicated parametrizations are not necessary in the considered regime.

\subsection{Normalization of QUBO instances}

Another important technical ingredient of the fpQAOA approach is the normalization of problem instances.
Consider a QUBO problem of the form~\eqref{eq:qubo_problem}, specified by an
$n\times n$ matrix $S$.
Multiplying the matrix by a positive constant $\kappa$,
\begin{equation}
    S \mapsto \kappa S,
\end{equation}
does not affect either the optimal solution or the feasible set
$\mathcal{F}(\alpha)$.
However, such a rescaling changes the dynamical phases accumulated during QAOA evolution.
Indeed, in order to preserve the same output statistics for the state
$\ket{\boldsymbol\beta,\boldsymbol\gamma}$,
all parameters $\gamma_i$ must be simultaneously transformed as
\begin{equation}
    \gamma_i \mapsto \kappa^{-1}\gamma_i.
\end{equation}

This issue becomes particularly important in the parameter-transfer setting,
where a single pair of fixed parameters
$(\boldsymbol\beta_*,\boldsymbol\gamma_*)$
is trained on an ensemble of problem instances and subsequently reused without modification during the sampling stage.
Without proper normalization, variations in the characteristic energy scale of different QUBO instances would lead to strongly misaligned phase dynamics and significantly reduce parameter transferability.

To mitigate this effect, each QUBO instance is rescaled according to
\begin{equation}
    S \mapsto
    S/\|S\|_{\mathrm{Frob}},
    \qquad
    \|S\|_{\mathrm{Frob}}
    =
    \Bigl(
        \sum_{i,j}\abs{s_{ij}}^2
    \Bigr)^{1/2}.
\end{equation}
Since
$\langle C(z) \rangle_z =2^{-n}{\rm Tr}H_C = 0$
and
$\expval{C^2(z)}_z =2^{-n}{\rm Tr}H_C^2 = \|S\|_{\mathrm{Frob}}^2$,
this transformation is equivalent to normalizing the standard deviation of the cost function
$C(z)$
over all computational basis states.
As a result, different problem instances acquire comparable characteristic phase scales during QAOA evolution, improving the consistency of the transferred parameters across the training ensemble.

Empirically, we observe that this normalization procedure works particularly well in combination with the AR-based objective and the two-parameter depth--size scaling ansatz introduced above.

\section{Performance demonstration} \label{sec:performance}
In this section, we evaluate the proposed fpQAOA scheme using numerical simulations performed with a GPU-accelerated statevector simulator of an ideal quantum processor.
As discussed previously, during the training stage the simulator is used to optimize the tuning parameters $u$ and $v$ in Eq.~\eqref{eq:fourier_2p}, thereby determining the fixed-point parameters $\boldsymbol\beta_*$ and $\boldsymbol\gamma_*$. During the sampling stage, the simulator is further employed to evaluate the resulting probabilities $P_{\alpha}(\boldsymbol\beta_*, \boldsymbol\gamma_*)$.

We emphasize that the use of a statevector simulator during the sampling stage is intended solely for benchmarking purposes. Such an approach would be infeasible on real quantum hardware for large-scale problems with high $n$, similarly to the fact that the exact optimal value of the cost function is generally unknown for hard combinatorial optimization problems.

As a benchmarking metric, we introduce the shots-to-solution (STS),
\begin{equation} \label{eq:sts}
    \mathrm{STS}_\alpha =
    P_\alpha(\boldsymbol\beta_*,\boldsymbol\gamma_*)^{-1},
\end{equation}
which provides the average number of measurement shots required to obtain the first bitstring belonging to ${\cal F}(\alpha)$, i.e., yielding a feasible solution with approximation ratio $\alpha$.
We note that the construction~\eqref{eq:sts} differs slightly from the commonly used time-to-solution (TTS) metric. In particular, TTS is additionally defined with respect to a target confidence probability of obtaining a desired solution, which should not be confused with the approximation-ratio threshold $\alpha$ used in our definition. The relation between STS and TTS is discussed in Appendix~\ref{app:sts_vs_tts}.

In this section, we consider two problem families:
(i) weighted MaxCut instances generated using the \textit{rudy} framework~\cite{rudy}, and
(ii) Sherrington--Kirkpatrick (SK) instances with normally distributed couplings.
In all cases, the problem matrices are normalized by their Frobenius norm, as discussed in the previous section.
Appendix~\ref{app:sm_more_problems} reports results for four additional QUBO families, including SK models on toric geometries and weighted MaxCut instances with exponential and Cauchy disorder.

For all problem families, the variational parameters $(u,v)$ are obtained using the random-mutations global optimization algorithm~\cite{chernyavskiy_randmut,chernyavskiy2013calculationquantumdiscordentanglement}.
The optimization is performed on a training set of $N_{\mathrm{train}}=200$ instances with $n=6$, and the objective is to maximize the minimum success probability $P_{0.95}$ [Eq.~\eqref{eq:succ_prob}] over the training set.
Within the sin--cos encoding~\eqref{eq:fourier_2p}, the optimization converges to
$u\approx 2.111$, $v\approx -0.468$ for weighted MaxCut, and
$u\approx -1.931$, $v\approx 0.424$ for SK instances.

\begin{figure*}[t]
    \centering
    \includegraphics[width=0.99\linewidth]{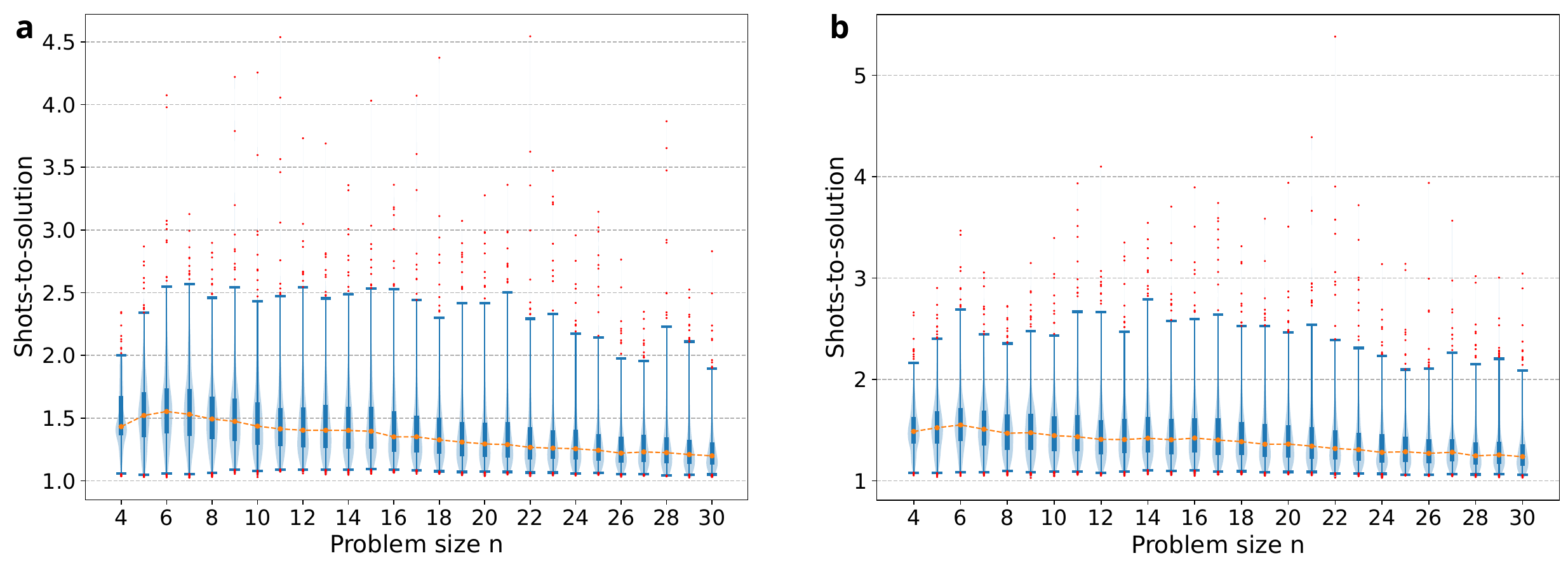}
    \caption{
    Empirical sampling complexity of fpQAOA trained on 200 $n=6$ instances with sin–-cos encoding, Frobenius normalization, and depth scaling $p=n$.
    For each problem size $n$, 1000 test instances are generated. 
    (a) Weighted MaxCut problem. 
    (b) Sherrington--Kirkpatrick Hamiltonian minimization problem. 
    Orange dashed line: median; blue boxes: Q1–Q3 interquartile range (25\%–75\%); ``error bars'': 1st and 99th percentiles; red dots: outliers.
    }
    \label{fig:main_res}
\end{figure*}

We then evaluate the resulting fixed-parameter ansatz on
$N_{\mathrm{test}} = 1000$
random QUBO instances for each system size
$n = 4,\ldots,30$.
Figures~\ref{fig:main_res}(a) and~\ref{fig:main_res}(b), corresponding to weighted MaxCut and SK instances, respectively, show that for $\alpha=0.95$ the median shots-to-solution does not increase with the problem size and, in fact, decreases as $n$ grows.
Remarkably, the same decreasing trend is observed even for the 99th percentile.
This behavior is the central empirical finding of our work: after training only on small instances, the fixed-parameter QAOA ansatz exhibits an apparently size-independent, and in this sense $O(1)$, sampling cost for finding high-quality approximate solutions within the studied range of problem sizes.

\begin{figure*}
    \centering
    \includegraphics[width=0.99\linewidth]{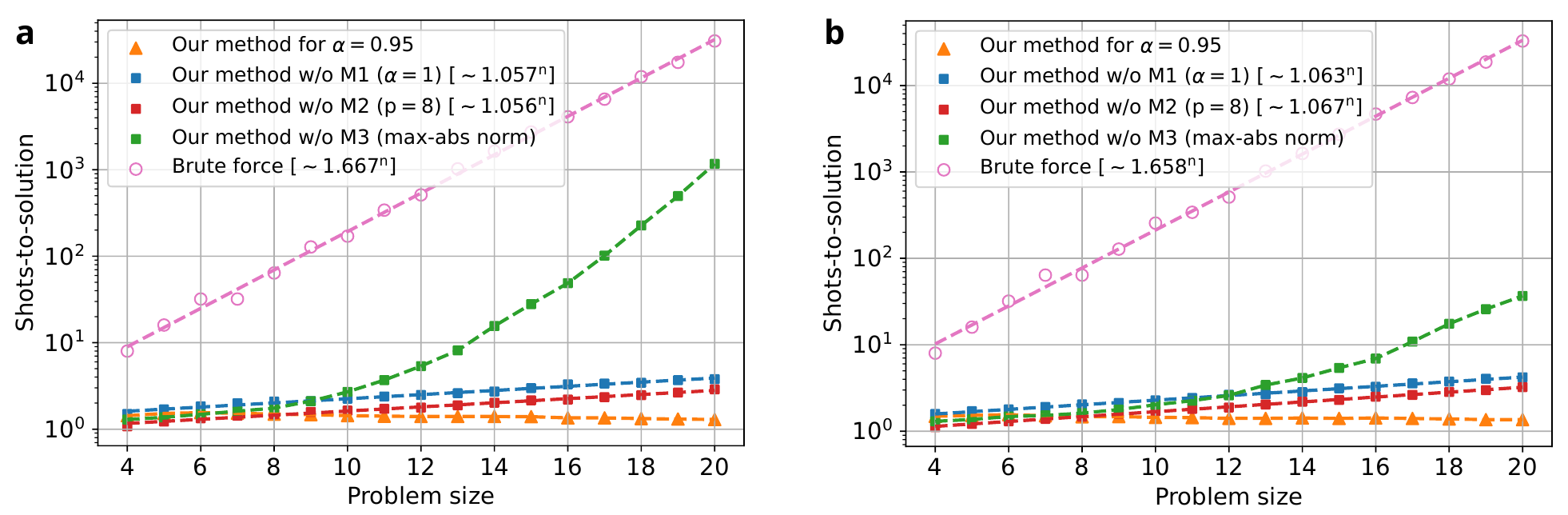}
    \caption{
    Comparison of the scaling of median STS values for fpQAOA at $\alpha=0.95$, 
    with all modifications enabled (same curves as in Fig.~\ref{fig:main_res}) and with one modification (M1, M2, or M3) removed. 
    The complexity of brute-force search is included as a baseline. 
    Exponential fits are shown in brackets when applicable. 
    (a) Weighted MaxCut problem.
    (b) Sherrington--Kirkpatrick Hamiltonian minimization problem.
    }
    \label{fig:fp-comparison}
\end{figure*}

One might argue that the observed behavior of the STS metric is merely a consequence of the increasing number of feasible solutions in ${\cal F}(\alpha)$, which naturally grows as $\alpha$ decreases.
To clarify the role of this effect, as well as the contribution of the introduced modifications M1--M3, we present the comparison shown in Fig.~\ref{fig:fp-comparison}. 

First, Fig.~\ref{fig:fp-comparison} shows the STS scaling for a brute-force random-sampling strategy at $\alpha=0.95$.
For weighted MaxCut and SK instances, respectively, the corresponding median STS scales approximately as $\sim 1.667^n$ and $\sim 1.658^n$.
This demonstrates that although the number of feasible solutions $|{\cal F}(\alpha)|$ grows exponentially with $n$ (reflected by an exponent smaller than $2$), their relative fraction within the full set of bitstrings still decreases exponentially with system size.
Therefore, the problem remains exponentially hard (in terms of direct search) even in the relaxed $\alpha<1$ setting.

Second, Fig.~\ref{fig:fp-comparison} illustrates the impact of removing individual components of the proposed approach.
Setting $\alpha=1$ (removing M1), fixing the circuit depth to $p=8$ (removing M2), or replacing Frobenius normalization with max-absolute-value scaling (removing M3; see also Appendix~\ref{sect:sm_diff_norms} for additional normalization strategies) all restore exponential growth of the median STS for both MaxCut and SK ensembles.
This behavior stands in sharp contrast to the approximately $O(1)$ scaling observed when all three modifications are combined.

\begin{figure*}
    \centering
    \includegraphics[width=0.99\linewidth]{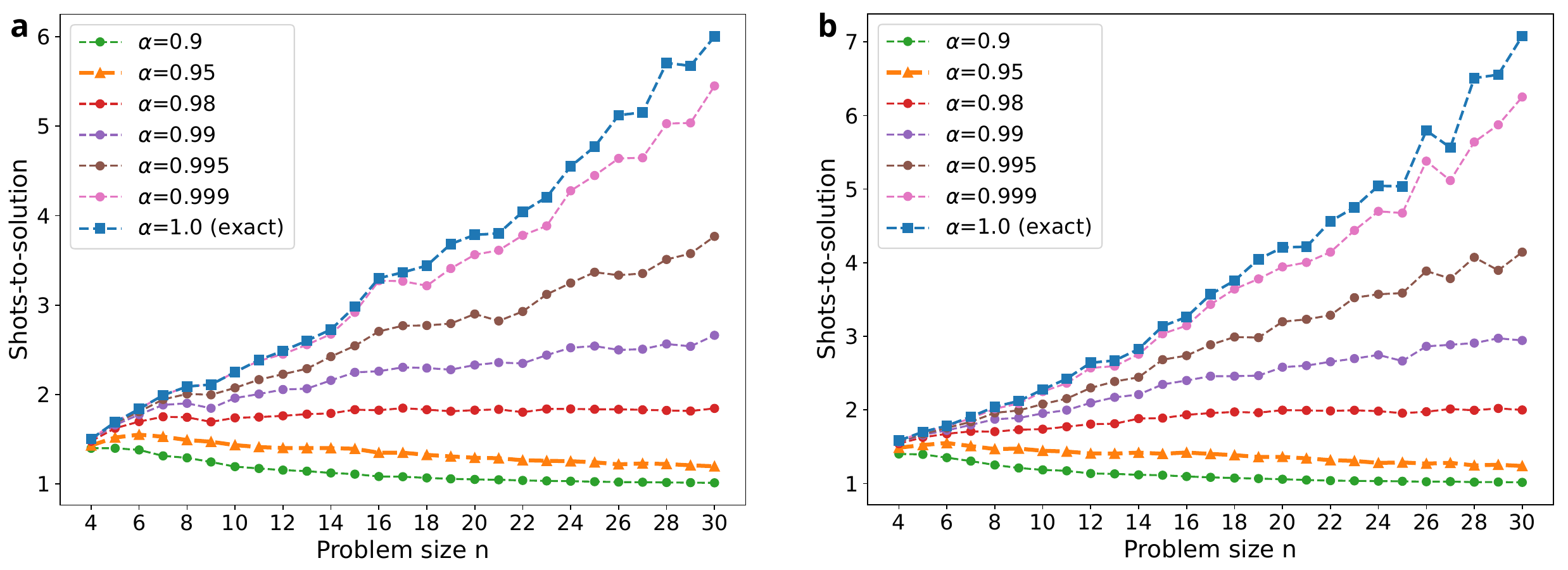}  
    \caption{Behavior of median STS values for different ARs $\alpha$, 
    including $\alpha=0.95$ (considered in Fig.~\ref{fig:main_res}) and $\alpha=1$ (considered in Fig.~\ref{fig:fp-comparison}). 
    (a) Weighted MaxCut problem. 
    (b) Sherrington--Kirkpatrick Hamiltonian minimization problem. 
    }
    \label{fig:fp-diff-ar}
\end{figure*}

Finally, Fig.~\ref{fig:fp-diff-ar} shows the dependence on the target approximation ratio.
For $\alpha\approx 1$, the STS scales exponentially, while for $\alpha=0.95$ it decreases in
all tested ensembles.  The transition between these regimes occurs when $1-\alpha$ reaches a
few percent, with slight variations depending on the problem distribution.  This behavior
aligns with the classical theory of probabilistic approximation algorithms~\cite{vazirani2001approximation}
and motivates studying adaptive thresholds $\alpha(n)$ that preserve non-increasing sampling
complexity.

\section{Discussion and outlook}
\label{sec:discussion}

In this work, we introduced a fpQAOA framework that combines
(i) approximation-ratio--oriented optimization,
(ii) depth scaling $p=n$ with a low-dimensional angle parameterization, and
(iii) Frobenius normalization of QUBO instances.

Extensive numerical simulations on more than $150\,000$ problem instances show that for approximation ratio $\alpha=0.95$ the median shots-to-solution (STS) does not increase with $n$ (up to $n=30$). Within the explored range of system sizes and problem classes, this is consistent with an effectively constant $O(1)$ median sampling complexity, while the time per shot scales polynomially because the circuit depth is $p=n$.

\BLUE{
Meanwhile, several important limitations and assumptions of the present study should be emphasized.

First, the proposed approach remains heuristic in nature.
Therefore, we do not claim a rigorous $O(1)$ complexity result for QAOA.
Instead, our conclusions are based on empirical scaling observed across large numerical datasets.
In particular, although the median STS remains approximately constant,
Fig.~\ref{fig:main_res} demonstrates the presence of nontrivial outliers with substantially
larger STS values.
This indicates that the observed scaling behavior should be interpreted as a statistical
property of the considered ensembles rather than a strict worst-case complexity guarantee.

Second, the present analysis is restricted to numerical simulations with system sizes up to
$n=30$ qubits.
While the observed trends remain stable throughout this regime,
it is currently unclear whether the same behavior persists asymptotically for larger systems.
Resolving this question likely requires experiments on larger quantum processors,
as well as the development of additional analytical tools for understanding the large-$n$
behavior of fixed-point QAOA protocols. 
However, we believe that these decreasing trends provide a more confident basis for extrapolation than increasing polynomial fits.

Another important limitation is the absence of noise in the current simulations of the sampling stage.
The present work assumes ideal coherent quantum evolution and therefore studies pure QAOA states
$\ket{\boldsymbol\beta,\boldsymbol\gamma}$.
On realistic quantum hardware, decoherence and gate imperfections transform the output state into
a mixed state that can be approximately represented as
\begin{equation}
    \rho(\boldsymbol{\beta}, \boldsymbol{\gamma})
    =
    F
    \ket{\boldsymbol\beta,\boldsymbol\gamma}
    \bra{\boldsymbol\beta,\boldsymbol\gamma}
    +
    (1-F)\rho_{\rm noise},
\end{equation}
where $F$ denotes an effective fidelity of the prepared state, while
$\rho_{\rm noise}$ represents the contribution of incoherent errors and
decoherence processes.
Within a simplistic noise model, one may
approximately treat $\rho_{\rm noise}$ as a maximally mixed state,
which corresponds to nearly random measurement outcomes and therefore
suppresses the probability of obtaining feasible solutions from
${\cal F}(\alpha)$.
Within the considered model, this probability scales approximately as $P_\alpha^{\rm noisy}\simeq F P_\alpha$,
which leads to a noisy version of the STS metric,
\begin{equation}
    {\rm STS}_{\alpha}^{\rm noisy}
    \simeq
    \frac{{\rm STS}_{\alpha}}{F}.
\end{equation}
Consequently, even moderate reductions of fidelity may significantly increase the required
sampling overhead.

This observation raises an important open question regarding the practical operating regime
of fpQAOA.
On one hand, the proposed protocol avoids the expensive classical outer-loop optimization,
which is advantageous for near-term quantum devices.
On the other hand, the scaling $p=n$ implies linearly growing circuit depth,
making the algorithm increasingly sensitive to decoherence and coherent control errors.
However, since STS decreases with $n$ there's still a potential for using increasing $p(n)$ that is smaller than $p=n$ (e.g. $p\propto\lfloor\sqrt{n}\rfloor$).
Finding the optimal trade‑off between depth and sampling complexity requires further investigation, so it remains unclear whether fpQAOA should be viewed primarily as a NISQ/EFTQC (early fault‑tolerant quantum~\cite{he2024parameter}) heuristic, or rather as a protocol better suited for future fault-tolerant quantum architectures.

Finally, our results indicate that the observed performance enhancement emerges from a
synergistic combination of several modifications rather than from any individual ingredient alone.
Removing AR targeting, normalization, or the fixed low-dimensional parameterization
restores exponential scaling of the shots-to-solution metric.
This suggests that scalable QAOA behavior may require coordinated adaptation of both the optimization
objective and the parameter-transfer strategy.
}

Overall, the presented results provide evidence that fixed-point parameter-transfer protocols
represent a promising direction for scalable quantum optimization.
Future work includes analytical investigation of the observed scaling laws,
extension to additional optimization ensembles and constraint structures,
study of noise resilience and error mitigation techniques,
and large-scale experimental implementation on quantum hardware.

\section*{Acknowledgments}

We thank S.~Usmanov for insightful discussions.
A part of the research is supported by the Priority 2030 program at the National University of Science and Technology ``MISIS'' under the project K1-2022-027. Extensive computations up to 30 qubits were supported by the state assignment of the National Research Center ``Kurchatov Institute''.

\section*{Data availability}
The data that support the findings of this article are openly
available~\cite{kulikov_2025_18099909}.

\appendix

\section{Results for different problem classes}\label{app:sm_more_problems}


Each problem class of QUBO instances~\eqref{eq:qubo_problem} has its own procedure for generating random coefficient matrices $S$.
Before applying the Frobenius normalization, we restrict ourselves to integer-valued matrices in order to ensure a numerically stable evaluation of the approximation ratio and the corresponding feasible set ${\cal F}(\alpha)$.
This becomes particularly important for large problem sizes, where the energy spectrum becomes increasingly dense and finite floating-point precision may lead to artificial misclassification of solutions near the approximation-ratio threshold.
In all considered problem ensembles, the diagonal elements are set to zero, i.e. $s_{ii}=0$.

In the main text, we presented the results for two problem classes, which are commonly used to evaluate QUBO solver performance:
\begin{itemize}
    \item Weighted MaxCut problem on random Erd\H{o}s--R\'enyi graphs with density $0.5$ and uniform distribution of integer weights from $-100$ to $100$ (MaxCut-Uniform): generated by Rudy framework \cite{rudy} with command \verb|rudy -rnd_graph $n 0.5 -random -100 100|.
    \item Sherrington–-Kirkpatrick Hamiltonian minimization problem (SK) with $s_{i<j}:=\lfloor 100000 \cdot \xi_{ij}\rceil$, where $\xi_{ij}\sim\text{Normal}(0,1)$.
\end{itemize}
In the latter case, just like in the Rudy framework, we multiply the random variable by a large number ($100000$ in our case) and then round it to the closest integer to make the problem integer-valued. In this section, we also analyze four more problem classes:
\begin{itemize}
    \item Weighted MaxCut problem on random Erd\H{o}s--R\'enyi graphs with density $0.5$ and exponential distribution of weights (MaxCut-Exp): $s_{i<j}=\lfloor 100000 \cdot \xi_{ij}\rceil \cdot M_{ij}$ with $\xi_{ij}\sim\text{Exp}(1)$.
    \item Weighted MaxCut problem on random Erd\H{o}s--R\'enyi graphs with density $0.5$ and Cauchy distribution of weights (MaxCut-Cauchy): $s_{i<j}=[100000 \cdot \xi_{ij}] \cdot M_{ij}$, where $\xi_{ij}\sim\text{Cauchy}(0, 1)$ and $M_{ij}$ is the binary mask indicating the presence of an edge in a graph, generated by \verb|networkx| module in Python.
    \item Sherrington–-Kirkpatrick Hamiltonian minimization on 2-rows toric graph (SK-2R): generated by Rudy framework \cite{rudy} with command \verb|rudy -spinglass2g 2 $columns| and $n/2$ columns.
    \item Sherrington–-Kirkpatrick Hamiltonian minimization on 3-rows toric graph (SK-3R): generated by Rudy framework \cite{rudy} with command \verb|rudy -spinglass2g 3 $columns| and $n/3$ columns.
\end{itemize}

\begin{figure*}[h]
    \centering
    \includegraphics[width=0.95\linewidth]{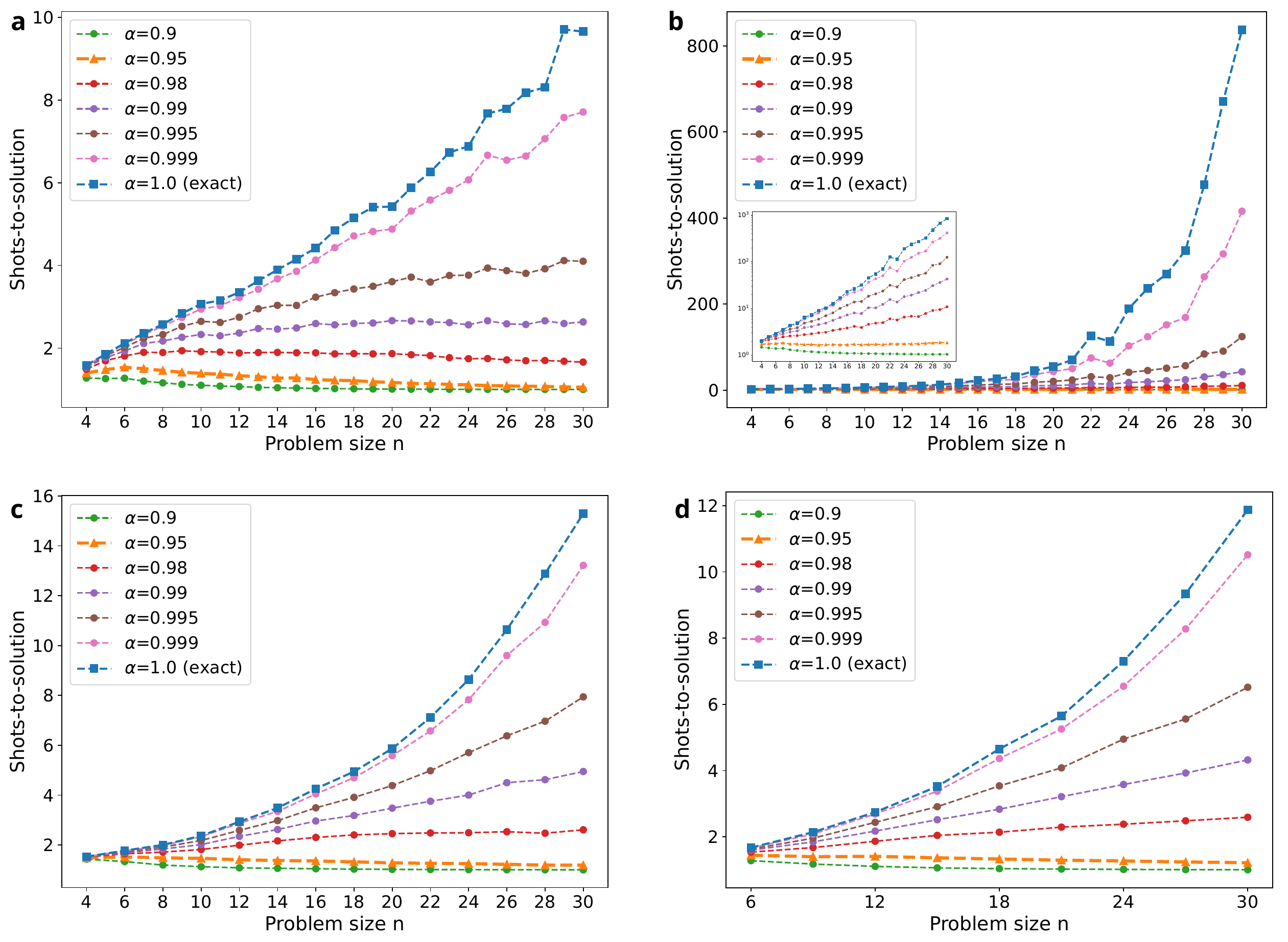}
    \caption{
    Behavior of median STS values for different ARs $\alpha$. 
    (a) Weighted MaxCut problem with exponential distribution of weights (MaxCut-Exp).
    (b) Weighted MaxCut problem with Cauchy distribution of weights (MaxCut-Cauchy); in the inset -- logarithmic scale along the y-axis.
    (c) Two-rows toric Sherrington--Kirkpatrick Hamiltonian minimization problem (SK-2R).
    (d) Three-rows toric Sherrington--Kirkpatrick Hamiltonian minimization problem (SK-3R).
    }
    \label{fig:ar_diff_problems}
\end{figure*}

In~Fig.~\ref{fig:ar_diff_problems}, we present the median STS metric versus the problem size for different values of AR $\alpha$. Similar to results in Sec.~\ref{sec:performance}, we observe the decreasing STS behavior for $\alpha \lesssim 0.98$. The exponential behavior of MaxCut-Cauchy is caused by the heavy-tailed Cauchy distribution: with increasing $n$ the probability to sample a high-valued graph weight is also increased, which introduces a high variability of function profiles. This, however, could be mitigated by instance-specific heuristic strategies.

\section{Comparison to other rescaling strategies}\label{sect:sm_diff_norms}

\begin{figure*}[h]
    \centering
    \includegraphics[width=0.95\linewidth]{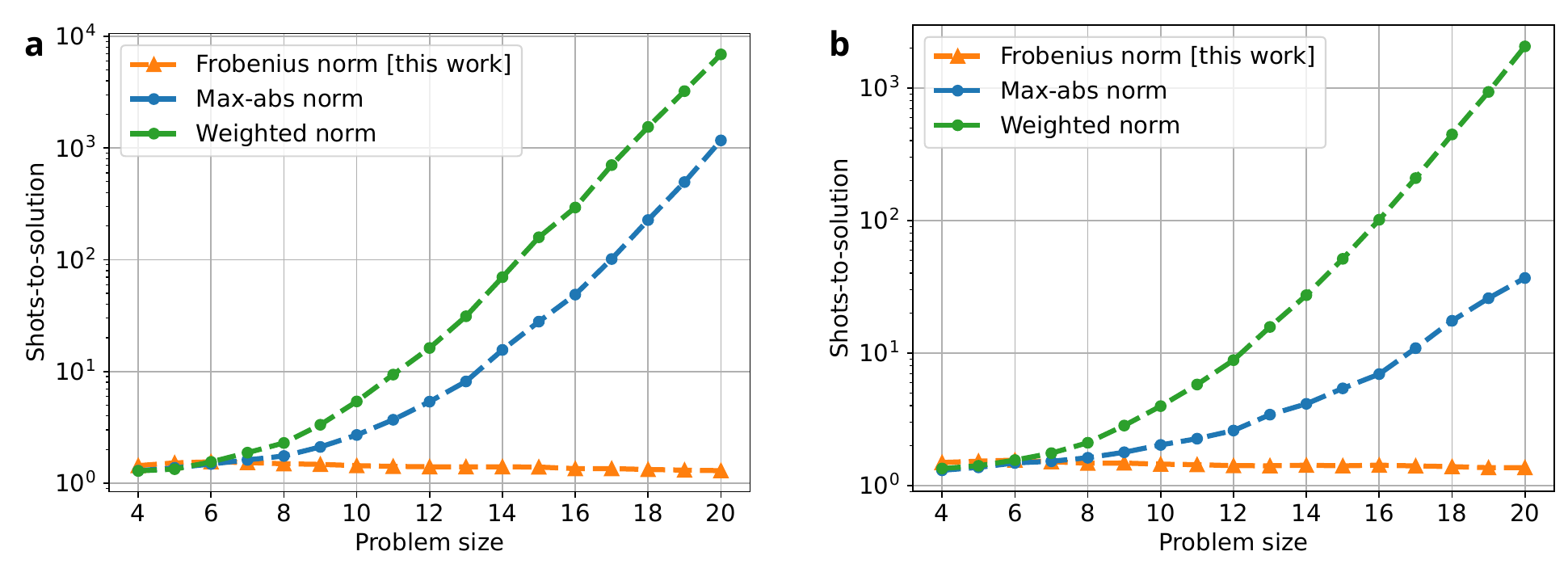}
    \caption{Comparison of different rescaling strategies and their influence on the median value of STS as a function of problem size $n$ (for AR $\alpha=0.95$). 
    All other settings of the fpQAOA approach are the same as in the main text.
    (a) Weighted MaxCut problem with uniform distribution of weights (MaxCut-Uniform).
    (b) Sherrington--Kirkpatrick Hamiltonian minimization problem (SK).}
    \label{fig:sm_diff_norms}
\end{figure*}

In our work, we rescale QUBO matrices [in the Ising form~\eqref{eq:qubo_problem}] so that their Frobenius norm equals one. 
For comparison, we also consider two other well-studied rescaling strategies from the literature: the max-abs approach~\cite{montanez2025toward,chernyavskiy2023fp,zalivako2025experimental}, defined as $\|S\|_{\rm max-abs}=\max_{ij}|s_{ij}|$, and the higher-order weighted-norm strategy~\cite{sureshbabu2024parameter},
\begin{equation}
    \|S\|_{\rm w.norm}=\sqrt{\frac{1}{|E_2|}\sum_{i<j}s_{ij}^2+\frac{1}{|E_1|}\sum_{i}s_{ii}^2},
\end{equation}
where $|E_k|$ with $k=1,2$ denotes the number of terms of order $k$. 
The results presented in Fig.~\ref{fig:sm_diff_norms} clearly show that these alternative normalization approaches lead to a rapid increase of the STS metric for our $p=n$ regime.

\BLUE{
\section{Relation to Adiabaticity}\label{app:adiabaticity}

QAOA can be viewed as a discretized version of adiabatic quantum optimization.
In the latter approach, the system is initialized in the ground state of the
initial Hamiltonian
$H_B=-\sum_{i=1}^n \sigma_x^{(i)}$
and evolves under a slowly varying Hamiltonian
\begin{equation}
    H(s)=(1-s)H_B+sH_C,
    \qquad s\in[0,1],
\end{equation}
where $s$ is the schedule parameter and $H_C$ is the cost Hamiltonian.
The final state can be approximated as
\begin{equation}\label{eq:adiabatic_state}
\begin{aligned}
    \ket{\Psi}&= {\cal T}
    \exp\left(-\imath\int_0^Tdt H[s(t)]\right)\ket{+}^{\otimes n}
    \\&\approx
    \prod_{i=1}^{p_{\rm ad}}e^{-\imath \Delta t_i H(s_i)}\ket{+}^{\otimes n}
    \\&\approx
    \prod_{i=1}^{p_{\rm ad}} e^{-\imath (1-s_i) \Delta t_i H_B}\, e^{-\imath s_i \Delta t_i H_C}
    \ket{+}^{\otimes n},
\end{aligned}
\end{equation}
where ${\cal T}$ denotes the time-ordering operator,
$T$ is the total evolution time, and the products are ordered from right to left with increasing index $i$.
The first approximation is obtained by replacing the continuous schedule
$s(t)$ with a piecewise-constant function consisting of
$p_{\rm ad}$ intervals of duration $\Delta t_i$.
The second approximation follows from the first-order Trotter--Suzuki formula.

The initial Hamiltonian $H_B$ differs by a sign from the conventional QAOA mixing Hamiltonian,
$H_M=\sum_{i=1}^n \sigma_x^{(i)}$, namely $H_B=-H_M$.
Therefore, evolution under $H_M$ with angle $\beta_i$ can equivalently be written as evolution under $H_B$ with angle
$\beta_i^+ := -\beta_i$.
Using this sign convention, the QAOA state~\eqref{eq:qaoa_state} can be rewritten as
\begin{equation}\label{eq:qaoa_state_hb}
\ket{\boldsymbol\beta,\boldsymbol\gamma}
=
\prod_{i=1}^p
e^{-\imath \beta_i^+ H_B},
e^{-\imath \gamma_i H_C}
\ket{+}^{\otimes n}.
\end{equation}

Comparing Eq.~\eqref{eq:qaoa_state_hb} with the first-order Trotterized adiabatic evolution in Eq.~\eqref{eq:adiabatic_state}, one can formally associate the QAOA angles with an effective discretized schedule by setting $p_{\rm ad}=p$ and
\begin{equation}
    \beta_i^+ = (1-s_i)\Delta t_i, \quad \gamma_i = s_i\Delta t_i.
\end{equation}
This gives
\begin{equation}
\Delta t_i = \beta_i^+ + \gamma_i,
\qquad
s_i=\frac{\gamma_i}{\beta_i^+ + \gamma_i}.
\end{equation}
Thus, each QAOA layer can be assigned an effective schedule parameter $s_i$ and an effective time step $\Delta t_i$.

Despite this formal correspondence, the fpQAOA trajectories considered in this work cannot be interpreted as a fine-grained Trotterization of adiabatic evolution. Indeed, for the optimized parameters, the effective time steps $\Delta t_i=\beta_i^+ + \gamma_i$
are not small, and therefore the assumptions underlying the standard adiabatic discretization are not satisfied.

Nevertheless, we observe that the resulting QAOA trajectory remains close to the low-energy region of the interpolating Hamiltonian family $H(s)$. This observation suggests that the fixed-point training procedure may implicitly learn an effective annealing schedule, despite operating outside the regime of conventional adiabatic evolution. Related questions may also be viewed in the broader context of counterdiabatic QAOA constructions~\cite{wurtz2022counterdiabaticity}, although we do not attempt to establish a direct connection here.

To illustrate this behavior, consider a random instance of the Sherrington--Kirkpatrick problem with $n=12$. Using the pretrained parameters $u\approx -1.931$ and $v\approx 0.424$ from the main text, we construct the corresponding QAOA angles $\boldsymbol\beta$ and $\boldsymbol\gamma$ using the sin--cos encoding with depth $p=n=12$.

Let $\ket{\psi_i}$ denote the state after the $i$th QAOA layer. For each layer and each value of the schedule parameter $s$, we compute the normalized energy
\begin{equation}\label{eq:norm_ev}
E(s;i)=
\frac{
\expval{H(s)}{\psi_i}
-
h_{\min}(s)}
{h_{\max}(s)-h_{\min}(s)},
\end{equation}
where $h_{\min}(s)$ and $h_{\max}(s)$ are the minimum and maximum eigenvalues of $H(s)$, respectively. Figure~\ref{fig:adiabaticity} shows that the effective schedule $s_i=\gamma_i / (\beta_i^+ + \gamma_i)$
associated with the fpQAOA parameters approximately tracks the low-energy region of the interpolating Hamiltonian family.

\begin{figure}[h]
    \centering
    \includegraphics[width=\linewidth]{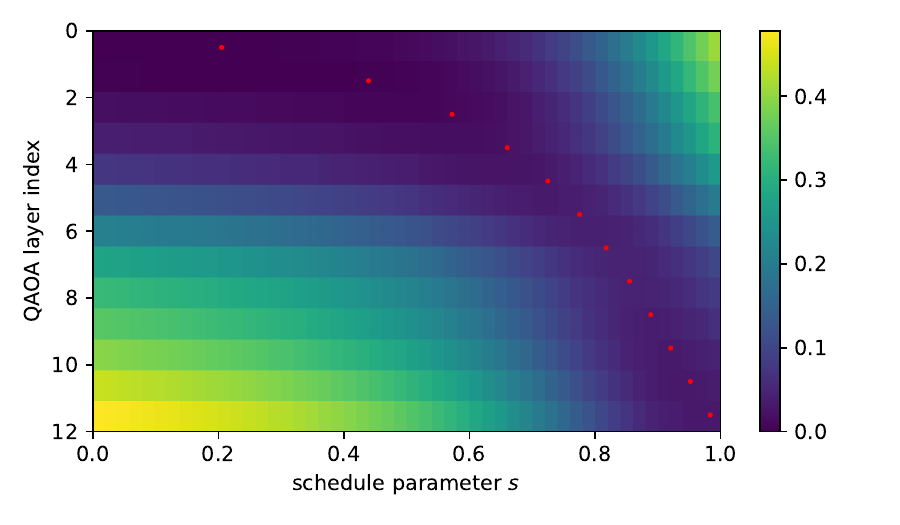}
    \caption{
        Normalized energy~\eqref{eq:norm_ev} of the intermediate QAOA states as a function of the effective schedule parameter $s$ for the interpolating Hamiltonian
        $H(s)=(1-s)H_B+sH_C$.
        Red dots correspond to the effective schedule
        $s_i=\gamma_i/(\beta_i^++\gamma_i)$
        associated with the fpQAOA parameters.
        The example shown corresponds to a Sherrington--Kirkpatrick instance with $n=12$ and QAOA depth $p=12$.
        }
    \label{fig:adiabaticity}
\end{figure}

}

\BLUE{
\section{Relation between STS and TTS} \label{app:sts_vs_tts}

To avoid terminological confusion, we note that in the quantum annealing and optimization literature, the term time-to-solution (TTS) is commonly used to denote the expected runtime required to achieve a target success probability $P_{\rm target}$:
\begin{equation}
    \mathrm{TTS}
    =
    t_{\rm shot}
    \frac{\ln(1-P_{\rm target})}{\ln(1-P_{\rm shot})},
\end{equation}
where $t_{\rm shot}$ is the time required for a single circuit execution (or annealing run), and $P_{\rm shot}$ is the probability of obtaining a desired solution in one shot.

In our case, the success probability is defined with respect to the approximation ratio threshold $\alpha$, such that $P_{\rm shot} \equiv P_\alpha$.
The metric used throughout the main text is the shots-to-solution (STS) quantity $\mathrm{STS}_\alpha = P_\alpha^{-1}$,
which corresponds to the average number of samples required to obtain the first solution from the feasible set ${\cal F}(\alpha)$.

Using the inequality
\begin{equation}
    \ln(1-x)\leq -x,
    \qquad x\in[0,1),
\end{equation}
one obtains the following upper bound:
\begin{equation}\label{eq:tts_sts}
    \mathrm{TTS}_{\alpha}
    \leq
    - t_{\rm shot}\cdot
    \ln(1-P_{\rm target})\cdot
    \mathrm{STS}_\alpha.
\end{equation}
For the commonly used choice $P_{\rm target}=0.99$, this relation becomes
\begin{equation}
    \mathrm{TTS}
    \lesssim
    4.6\, t_{\rm shot}\, \mathrm{STS}_\alpha.
\end{equation}

Therefore, STS and TTS exhibit the same monotonic dependence on the success probability $P_\alpha$, and become asymptotically equivalent up to a constant prefactor when $P_\alpha \ll 1$.

Traditionally, TTS is defined for obtaining an exact optimal solution, whereas approximation-ratio-based metrics quantify the probability of obtaining sufficiently good approximate solutions without explicitly accounting for runtime~\cite{pelofske2025comparing}. In this work, we combine these two perspectives by introducing STS$_\alpha$, which characterizes the expected sampling complexity for obtaining approximate solutions with a prescribed approximation ratio.
}

\bibliographystyle{apsrev4-1}
\bibliography{bibliography}

\end{document}